\documentclass[prb,showpacs,
twocolumn, aps]{revtex4}

\usepackage{graphicx}
\usepackage{dcolumn}
\usepackage{bm}

\begin{document}

\title{Photoemission study of polycrystalline Gd$_{1-x}$Sr$_x$TiO$_{3+\delta}$: correlation and surface effects}

\author{M. Sing, M. Karlsson, D. Schrupp, R. Claessen, M. Heinrich, V. Fritsch,
H.-A. Krug von Nidda, A. Loidl, and R. Bulla}
\affiliation{Institut f\"ur Physik, Universit\"at Augsburg,
D-86135 Augsburg, Germany}

\date{\today}

\begin{abstract}
We report photoemission studies of polycrystalline samples of
Sr-doped GdTiO$_{3+\delta}$, which undergoes a Mott-Hubbard-like
metal-insulator transition at a Ti-concentration of $\approx
20$\%. The Ti $3d$-derived valence band near the Fermi level
displays a two-peak structure consistent with a Hubbard-model-type
interpretation as quasiparticle and lower Hubbard band. Contrary
to the theoretical expectation the lower Hubbard band does not
change its energy with doping. However, the analysis of the core
level spectra and elemental mapping with a scanning electron
microscope indicate a strongly inhomogeneous doping concentration
in the probed surface regions. This questions the intrinsic
character of the valence band spectra and demonstrates a general
difficulty when using polycrystalline samples for such studies.

\end{abstract}


\maketitle

Many transition metal oxides display metal-insulator transitions
(MITs) upon variation of external parameters such as temperature,
pressure, or chemical doping. These result from a competition
between local Coulomb interactions in the open metal $3d$ shell
and electron delocalization induced by hopping via the oxygen
ligands. \cite{Imada98} A particularly simple class of systems are
the perovskite-like titanates $R$TiO$_3$, where $R$ denotes a
trivalent rare earth ion such as Y$^{3+}$ or La$^{3+}$. This
leaves the Ti-ion in a $3d^1$ state and should according to band
theory result in a metallic ground state. However, the strong
on-site $3d$ Coulomb energy turns the undoped material into a
Mott-insulator.\cite{Imada98} Replacing the trivalent $R$-ions by
divalent ions reduces the Ti~$3d$ occupation from an integer value
and eventually restores the metallic state beyond a critical
doping level. Such MITs have been investigated in detail for,
e.g., La$_{1-x}$Sr$_x$TiO$_3$ and Y$_{1-x}$Ca$_x$TiO$_3$.
\cite{Tokura93,Taguchi93}

The physics of the Mott transition is theoretically best described
within the Hubbard model framework. Recently, enormous progress
has been made in its treatment to allow detailed calculations of
physical properties which can be compared to
experiment.\cite{Imada98,Georges96,Gebhard97,Held01} The most
direct test of the theory concerns its spectral properties. For
example, the single-particle excitation spectrum (or rather its
electron removal part) can in principle be probed by photoelectron
spectroscopy.\cite{Huefner95} Photoemission studies on early
transition metal perovskites with doping-induced MITs
\cite{Fujimori92a,Fujimori92b,Morikawa96,Pen99,Maiti00,Fujimori01}
indeed confirm some of the predictions of the Hubbard model. For
example, they observe a characteristic double-peak structure of
the Ti $3d$ spectrum which is consistent with the expected
splitting into a quasiparticle peak and an incoherent lower
Hubbard band. On the other hand, there are also qualitative
discrepancies between theory and photoemission, concerning
especially the effect of doping on the energy of these spectral
features. One has to keep in mind though that photoemission is an
extremely surface sensitive method and that these studies have
been performed on scraped surfaces of polycrystalline material.
Very little is known about quality, atomic geometry, defect
density, or homogeneity of these surfaces.

Here we present photoemission results on
Gd$_{1-x}$Sr$_x$TiO$_{3+\delta}$ which also displays a
doping-induced MIT.\cite{Heinrich01} The Ti $3d$ spectra are
similar to those observed in other titanates and display overall
qualitative agreement with calculated Hubbard model spectra.
However, from peculiar charging effects in the core level spectra
we find that the doping concentration in the probed surface layer
is non-homogeneous. Chemical microananalysis by scanning electron
microscopy confirms this result. These findings invalidate any
interpretation of the valence band spectra as those of a
single-phased compound and question the use of polycrystalline
material for such studies.

The low energy properties of Gd$_{1-x}$Sr$_x$TiO$_{3+\delta}$ have
been studied in detail in Ref.~\onlinecite{Heinrich01}. Undoped
GdTiO$_3$ is a Mott insulator which at low temperatures undergoes
ferrimagnetic ordering. Hole-doping by substituting Sr for Gd
leads to an insulator-metal transition at $x \approx 0.2$. In the
metallic regime the low energy properties are observed to follow
Fermi liquid-type behavior at least up to $x=0.5$. Approaching the
Mott transition from the metallic side the Sommerfeld coefficient
of the specific heat increases up to 50 times with respect to
simple metals. This indicates a strong enhancement of the
effective carrier mass and thus the formation of heavy fermion
quasiparticles.\cite{Heinrich01} Upon further doping, a second MIT
occurs slightly above $x=0.7$ concomitant with a structural change
from orthorombic to cubic symmetry. Pure SrTiO$_3$ is a band
insulator.

Ceramic samples of Gd$_{1-x}$Sr$_x$TiO$_{3+\delta}$ have been calcinated
from TiO$_2$, SrCO$_3$, Gd$_2$O$_3$ and Ti$_2$O$_3$ high purity
powders and pressed into pellets. These pellets have been annealed at
$1200^{\circ}$C under N$_2$ atmosphere for 15 hours and finally arc-melted
under argon atmosphere. From thermogravimetric measurements the
amount of excess oxygen was found to be $\delta \approx 0.3$, except
for $x=0.6$ and $0.8$ where $\delta = 0.12$ and $0.06$, respectively.
Assuming formal valencies of Gd$^{3+}$, Sr$^{2+}$, and O$^{2-}$ the
nominal occupation of the Ti $3d$ shell is given by $n_{3d} =
1-x-2\delta$. The samples used here are identical to the ones
studied in Ref.~\onlinecite{Heinrich01}.

Valence band spectra were obtained by ultraviolet photoelectron
spectroscopy (UPS) using He I radiation (21.2 eV). For x-ray
photoemission (XPS) of the core levels we used monochromatized Al
K$_{\alpha}$ radiation (1486.6 eV). The energy resolution was 45
meV and 1.3 eV for UPS and XPS, respectively. Clean surfaces of
the polycrystalline samples were prepared by {\it in situ}
scraping with a diamond file. The cleanliness of the surfaces was
characterized by the absence of a C $1s$ signal and by a
well-defined O $1s$ peak.

\begin{figure}[b]
\begin{center}\mbox{}
\includegraphics[width=0.45\textwidth]{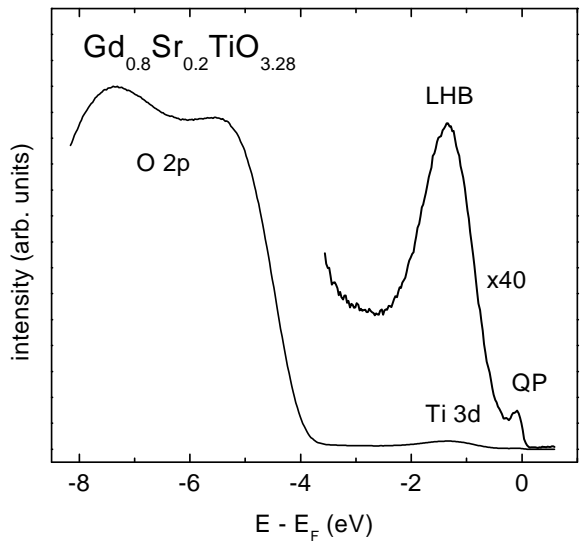}
\end{center}
\caption{UPS valence band spectrum of
Gd$_{0.8}$Sr$_{0.2}$TiO$_{3.28}$. The spectrum has been corrected
for He I$_{\beta}$ satellite emission. The Ti $3d$-derived part of
the spectrum decomposes into the lower Hubbard band (LHB) and a
quasiparticle peak (QP) at the Fermi level.}\label{valenceband}
\end{figure}

A typical UPS valence band spectrum of Gd$_{1-x}$Sr$_x$TiO$_{3+\delta}$
is shown in Fig.~\ref{valenceband}. The intense feature below
-4 eV is formed by bands of predominantly O $2p$ character. The small
emission closer to the Fermi level is attributed to Ti $3d$
states.\cite{Fujimori92b} Its lineshape displays a characteristic
double structure with a broad feature centred at -1.3 eV and a sharp
much less intense peak at the Fermi level. By analogy to the
valence band spectra of other titanates
\cite{Fujimori92a,Fujimori92b,Morikawa96,Fujimori01}
and by comparison to theoretical spectra
(discussed below) these structures are identified as lower Hubbard band (LHB) and
quasiparticle band (QP), respectively.

\begin{figure}
\begin{center}\mbox{}
\includegraphics[width=0.45\textwidth]{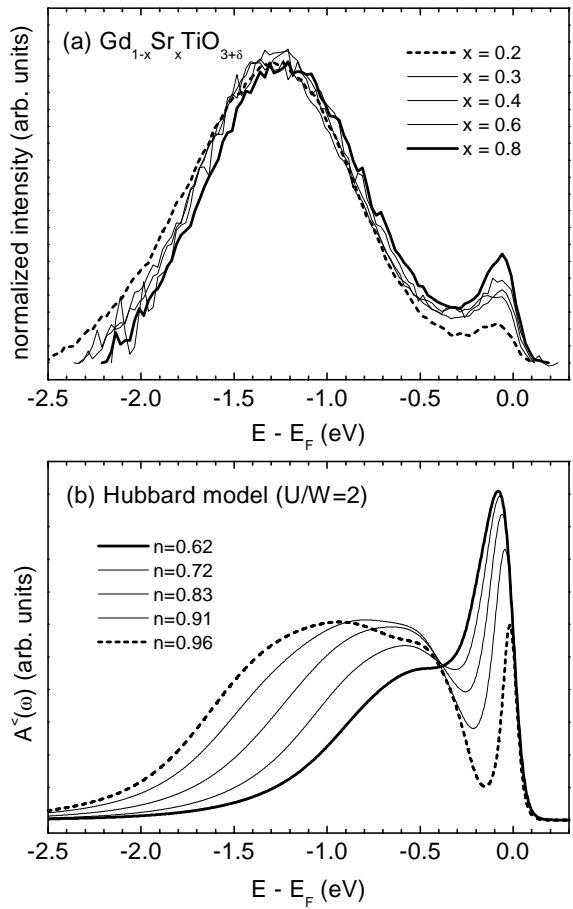}
\end{center}
\caption{(a) Ti $3d$ spectra of Gd$_{1-x}$Sr$_{x}$TiO$_{3+\delta}$
with varying Sr-doping. For better clarity an inelastic background is
subtracted. The spectra are normalized to same integral Ti $3d$
weight.
(b) Theoretical photoemission spectra $A^{<}(\omega)$ calculated for the Hubbard model
at various band filling factors $n$. The ratio of on-site Coulomb energy $U$ and band width $W$
is chosen as $U/W=2$, with W set to 2 eV. See text for further details.}\label{experiment-vs-theory}
\end{figure}

The evolution of the Ti $3d$ spectrum with Sr-doping is shown in
Fig.~\ref{experiment-vs-theory}(a). In order to facilitate better
comparison the spectra have been normalized to same integral area; without
this normalization the spectral Ti $3d$ weight decreases
relative to the O $2p$ signal, in line with the reduced $3d$
occupancy. All spectra display the characteristic decomposition into
a LHB peak and a quasiparticle band. The latter is most
intense for high doping ($x=0.8$, but still within the metallic regime of
the phase diagram) and strongly looses its spectral weight relative
to the LHB as one approaches the Mott transition.

In Fig.~\ref{experiment-vs-theory}(b) we display the electron
removal part of the spectral function of the single-band Hubbard
model as function of the band filling $n$, calculated within
dynamical mean-field theory (DMFT) \cite{Georges96,Metzner89}
using the numerical renormalization group (NRG)
method.\cite{Bulla-Sammelref} The theoretical spectra have been
broadened by the experimental resolution. They nicely reproduce
the two peak structure of the measured UPS spectra. Another
qualitative agreement between theory and experiment is the gain of
quasiparticle weight for increased hole-doping. However, there is
also an important qualitative discrepancy concerning the doping
dependence of the peak {\it energies}. In the Hubbard model the
quasiparticle peak shifts into the LHB with increased doping, or
rather, with all energies referenced to the Fermi level $E_F$, the
LHB peak moves closer to the Fermi level. In the theoretical
spectra of Fig.~\ref{experiment-vs-theory}(b) this shift amounts
to almost 1 eV from $n=0.96$ to 0.62. In contrast, in our UPS
spectra the energy position of the LHB peak is completely
independent of doping, consistent with photoemission results
obtained on other doped
titanates.\cite{Fujimori92b,Morikawa96,Fujimori01} One may wonder,
if this discrepancy is an artefact of the single-band Hubbard
model used here, whereas in the perovskite titanates the Ti $3d$
electrons have an added orbital degree of freedom as they move in
a $t_{2g}$ triplet of bands. There is however theoretical
indication that the doping-dependent relative energy shift of LHB
and quasiparticle peak is preserved in multiband Hubbard
models.\cite{Bulla-Comment}

\begin{figure}
\begin{center}\mbox{}
\includegraphics[width=0.45\textwidth]{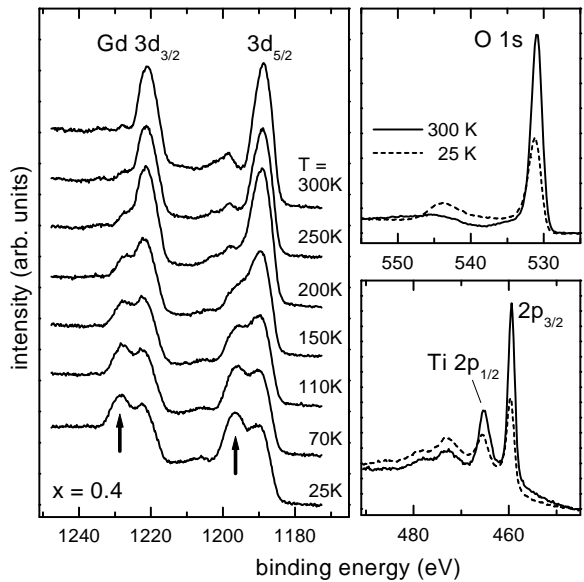}
\end{center}
\caption{XPS core level spectra of Gd$_{0.6}$Sr$_{0.4}$TiO$_{3.28}$.
Left panel: Temperature-dependence of the Gd $3d$ spectra. Note
the evolution of an additional structure (denoted by arrows) below
200 K. Right panels: O $1s$ and Ti $2p$ spectra of the same sample at
room temperature (solid line) and 25 K (dashed line).}\label{T-dependence}
\end{figure}

Rather, the observed disagreement may originate in an
inhomogeneous sample composition which we infer from a detailed
analysis of our XPS core level spectra. As an example,
Fig.~\ref{T-dependence} shows the Gd $3d$ spectrum of a $x=0.4$
sample, measured as function of temperature between 300 and 25 K.
At room-temperature we observe the clean multiplet structure of
trivalent Gd.\cite{Lademan96} Upon lowering the temperature
(starting between 200 and 150 K) a gradual shift of spectral
weight to higher binding energy occurs until at the lowest
temperatures additional structures have developed (denoted by
arrows in Fig.~\ref{T-dependence}). A detailed analysis of the low
temperature lineshape indicates that it can largely be understood
as a superposition of the original room-temperature spectrum and
an identical replica shifted by $\sim 9$ eV to higher energy.
Temperature-dependent line splitting and concomitant energy shifts
were found for {\it all} core levels. As examples
Fig.~\ref{T-dependence} also contains the O $1s$ and Ti $2p$
spectra.\cite{Ti-2p-comment} Furthermore, the effect was observed
for all samples studied here except for $x=0.8$.

\begin{figure}[b]
\begin{center}
\includegraphics[width=0.45\textwidth]{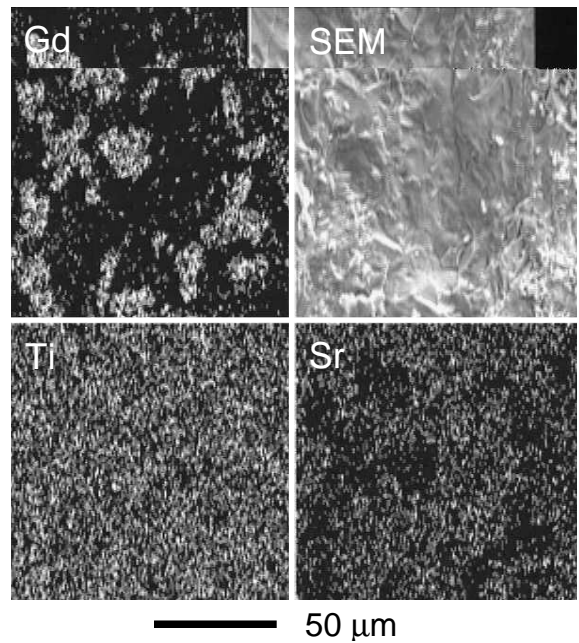}
\end{center}
\caption{Scanning electron micrograph of a scraped surface
of polycrystalline Gd$_{0.7}$Sr$_{0.3}$TiO$_{3.3}$,
measured with a primary electron energy of 20 keV (upper right panel).
Also shown are the elemental maps for Gd, Ti, and Sr simultaneously obtained
by EDX. White (black) areas denote high (low) concentration.}\label{EDX-SEM}
\end{figure}

The peculiar temperature dependence of the core level spectra
can be explained in a quite natural way, if we assume an inhomogeneous doping
concentration, {\it i.e.} a decomposition of our ceramic sample
into metallic and insulating domains. In this case the doping parameter $x$ represents
only the average composition. At low temperatures the insulating fraction
cannot compensate the photo-induced loss of electrons and will
charge up, thereby leading to an apparent shift of the (partial) spectra to higher
binding energy. With increased temperature thermally activated charge
carriers help to reduce the charging in the non-metallic parts
until eventually their core level energies approach those of the
metallic phases. The absence of charging effects at a nominal doping
of $x=0.8$ indicates that this sample consists only of metallic
phases (though it may still be inhomogeneous) and that the insulating
fractions at lower $x$ belong to the Gd-rich Mott-Hubbard
insulator phase.

The chemical phase separation suggested by the XPS data is confirmed by
energy dispersive x-ray microanalyis (EDX) using a scanning electron
microscope. Figure \ref{EDX-SEM} shows elemental maps taken on the
$x=0.3$ (nominal value) sample. The Gd signal displays a highly
inhomogeneous distribution which anticorrelates with the Sr
density, {\it i.e.} Gd-rich regions coincide with Sr-poor ones and
{\it vice versa}. In comparison, Ti appears to be rather evenly
distributed over the sampled area. This gives independent evidence that
the doping concentration strongly fluctuates inducing a spatial
segregation into metallic and (Mott) insulating phases. Similar
EDX maps were obtained from our other samples.

The fact that the spatial phase separation observed here has gone
undetected in measurements of the bulk properties of the same
samples \cite{Heinrich01} suggests that it occurs only on the
grain surfaces of the polycrystalline material. The thickness of
the inhomogeneous surface layer must be at least of the order of
the XPS probing depth ($30...50$\,{\AA}). The observation of
inhomogeneities by EDX (sampling depth $\sim 1 \mu$m) indicates
that it is probably much larger.

As a consequence of these findings we must also conclude that the
UPS valence band spectra presented above {\it do not} reflect the intrinsic
electronic structure of single-phased material. They rather represent
superpositions of spectra corresponding to the individual doping profile
of each sample. Even though we did observe trends in the Ti $3d$ spectra
with nominal doping $x$, the true doping dependence of the intrinsic spectra will
thus be strongly obscured. This may be one possible explanation for the
observed discrepancy to the Hubbard model, namely for the absence of
relative energy shifts between LHB and quasiparticle band in our UPS spectra.

As mentioned before, our valence band data on
Gd$_{1-x}$Sr$_{x}$TiO$_{3+\delta}$ reproduce those obtained from
other doped titanates.\cite{Fujimori92b,Morikawa96,Fujimori01} As
these studies also employed polycrystalline samples, they may
suffer from the same problems encountered here. Unfortunately,
they do not give sufficient information in order to allow an
assessment of their sample homogeneity. This implies that the
photoemission data on the doping effects in these compounds
presently available in the literature may not be of sufficient
quality to allow a detailed and reliable comparison to theoretical
spectra. Based on our present findings we strongly advocate the
use of {\it single crystals} for further photoemission studies of
the doping-induced MIT in titanates and related perovskites,
because for such samples it is much easier to ensure and
characterize sample homogeneity. We emphasize that this does not
automatically imply that the surface electronic structure probed
by photoemission is identical to that of the volume. In fact,
recent studies on vanadates indicate substantial differences
between surface and bulk.\cite{Maiti01,Suga02}

In summary, we have presented the first photoemission data on the
doping dependence of Gd$_{1-x}$Sr$_{x}$TiO$_{3+\delta}$. The Ti $3d$ part of
the valence band is found to be in partial qualitative agreement with Hubbard
model calculations, but also displays notable discrepancies to
theory. From our XPS and EDX results we conclude that the surfaces of the
polycrystalline samples are chemically inhomogeneous, thereby
questioning the intrinisic character of the valence band spectra.
Further studies require the use of high-quality single crystals.

We thank O.S.~Becker for the thermogravimetric characterization of the samples.
We also gratefully acknowledge support by the Deutsche Forschungsgemeinschaft
through the Sonderforschungsbereich 484.


\begin{thebibliography}{10}
\bibitem{Ti-2p-comment}
While for O $1s$ the effect is clearly visible, the Ti $2p$
replica appearing at
  low temperature interfere with relatively strong plasmon satellites already
  existing at room-temperature.
\end{thebibliography}
\end{document}